\documentclass[11pt]{article}
\usepackage{amssymb}
\usepackage{geometry}
\usepackage{amsmath}
\usepackage{amsfonts}
\usepackage{graphicx}

\begin{document}
\newcommand{\2}{\vspace{0.2 cm}}
\newcommand{\dist}{{\rm dist}}
\newcommand{\diam}{{\rm diam}}
\newcommand{\rad}{{\rm rad}}
\newcommand{\dom}{\mbox{$\rightarrow$}}
\newcommand{\ndom}{\mbox{$\not\rightarrow$}}
\newcommand{\sdom}{\mbox{$\Rightarrow$}}
\newcommand{\nsdom}{\mbox{$\not\Rightarrow$}}
\newcommand{\qed}{\hfill$\diamond$}
\newcommand{\pf}{{\bf Proof: }}
\newtheorem{theorem}{Theorem}[section]
\newcommand{\ra}{\rangle}
\newcommand{\la}{\langle}
\newtheorem{lemma}[theorem]{Lemma}
\newtheorem{corollary}[theorem]{Corollary}
\newtheorem{proposition}[theorem]{Proposition}
\newtheorem{conjecture}[theorem]{Conjecture}
\newtheorem{problem}[theorem]{Problem}
\newtheorem{remark}[theorem]{Remark}
\newtheorem{example}[theorem]{Example}
\newcommand{\beq}{\begin{equation}}
\newcommand{\eeq}{\end{equation}}
\newcommand{\argmax}{{\rm argmax}}
\newcommand{\MiP}{MinHOMP($H$) }
\newcommand{\MaP}{MaxHOMP($H$) }
\newcommand{\vecc}[1]{\stackrel{\leftrightarrow}{#1}}

\title{Minimum Cost Homomorphisms to Semicomplete Multipartite Digraphs}
\author{Gregory Gutin\thanks{Corresponding author. Department of Computer Science,
Royal Holloway University of London, Egham, Surrey TW20 OEX, UK,
gutin@cs.rhul.ac.uk and Department of Computer Science, University
of Haifa, Israel} \and Arash Rafiey\thanks{Department of Computer
Science, Royal Holloway University of London, Egham, Surrey TW20
OEX, UK, arash@cs.rhul.ac.uk} \and Anders Yeo\thanks{Department of
Computer Science, Royal Holloway University of London, Egham,
Surrey TW20 OEX, UK, anders@cs.rhul.ac.uk}}

\date{}

\maketitle

\begin{abstract}
For digraphs $D$ and $H$, a mapping $f:\ V(D)\dom V(H)$ is a {\em
homomorphism of $D$ to $H$} if $uv\in A(D)$ implies $f(u)f(v)\in
A(H).$ For a fixed directed or undirected graph $H$ and an input
graph $D$, the problem of verifying whether there exists a
homomorphism of $D$ to $H$ has been studied in a large number of
papers. We study an optimization version of this decision problem.
Our optimization problem is motivated by a real-world problem in
defence logistics and was introduced very recently by the authors
and M. Tso.

Suppose we are given a pair of digraphs $D,H$ and a positive
integral cost $c_i(u)$ for each $u\in V(D)$ and $i\in V(H)$. The
cost of a homomorphism $f$ of $D$ to $H$ is $\sum_{u\in
V(D)}c_{f(u)}(u)$. Let $H$ be a fixed digraph. The  minimum cost
homomorphism problem for $H$, MinHOMP($H$), is stated as follows:
For input digraph $D$ and costs $c_i(u)$ for each $u\in V(D)$ and
$i\in V(H)$, verify whether there is a homomorphism of $D$ to $H$
and, if it does exist, find such a homomorphism of minimum cost.
In our previous paper we obtained a dichotomy classification of
the time complexity of \MiP for $H$ being a semicomplete digraph.
In this paper we extend the classification to semicomplete
$k$-partite digraphs, $k\ge 3$, and obtain such a classification
for bipartite tournaments.
\end{abstract}

\section{Introduction}

In our terminology and notation, we follow \cite{bang2000}. In
this paper, directed (undirected) graphs have no parallel arcs
(edges) or loops. The vertex (arc) set of a digraph $G$ is denoted
by $V(G)$ ($A(G)$). The vertex (edge) set of an undirected graph
$G$ is denoted by $V(G)$ ($E(G)$). A digraph $D$ obtained from a
complete $k$-partite (undirected) graph $G$ by replacing every
edge $xy$ of $G$ with arc $xy$, arc $yx$, or both $xy$ and $yx$,
is called a {\em semicomplete $k$-partite digraph} (or, {\em
semicomplete multipartite digraph} when $k$ is immaterial). The
{\em partite sets} of $D$ are the partite sets of $G.$ A
semicomplete $k$-partite digraph $D$ is {\em semicomplete} if each
partite set of $D$ consists of a unique vertex. A {\em $k$-partite
tournament} is a semicomplete $k$-partite digraph with no directed
cycle of length 2. Semicomplete $k$-partite digraphs and its
subclasses mentioned above are well-studied in graph theory and
algorithms, see, e.g., \cite{bang2000}.

For excellent introductions to homomorphisms in directed and
undirected graphs, see \cite{hell2003,hell2004}. For digraphs $D$
and $H$, a mapping $f:\ V(D)\dom V(H)$ is a {\em homomorphism of
$D$ to $H$} if $uv\in A(D)$ implies $f(u)f(v)\in A(H).$ A
homomorphism $f$ of $D$ to $H$ is also called an {\em
$H$-coloring} of $G$, and $f(x)$ is called a {\em color} of $x$
for every $x\in V(D).$ We denote the set of all homomorphisms from
$D$ to $H$ by $HOM(D,H)$.

For a fixed digraph $H$, the {\em homomorphism problem} ${\rm
HOMP}(H)$ is to verify whether, for an input digraph $D$, there is
a homomorphism of $D$ to $H$ (i.e., whether $HOM(D,H)\neq
\emptyset$). The problem ${\rm HOMP}(H)$ has been studied for
several families of directed and undirected graphs $H$, see, e.g.,
\cite{hell2003,hell2004}. The well-known result of Hell and
Ne\v{s}et\v{r}il \cite{hellJCT48} asserts that ${\rm HOMP}(H)$ for
undirected graphs is polynomial time solvable if $H$ is  bipartite
and it is NP-complete, otherwise. Such a dichotomy classification
for all digraphs is unknown and only partial classifications have
been obtained; see \cite{hell2004}. For example, Bang-Jensen, Hell
and MacGillivray \cite{bangSIAMJDM1} showed that ${\rm HOMP}(H)$
for semicomplete digraphs $H$ is polynomial time solvable if $H$
has at most one cycle and ${\rm HOMP}(H)$ is NP-complete,
otherwise.

The authors of \cite{gutinDAM} introduced an optimization problem
on $H$-colorings for undirected graphs $H$, \MiP (defined below).
The problem is motivated by a problem in defence logistics. In our
previous paper \cite{gutinsd}, we obtained a dichotomy
classification for the time complexity of \MiP for $H$ being a
semicomplete digraph. In this paper, we extend that classification
to obtain a dichotomy classification for semicomplete $k$-partite
digraphs $H$, $k\ge 3$. We also obtain a classification of the
complexity of \MiP when $H$ is a bipartite tournament. The case of
arbitrary semicomplete bipartite digraphs appears to be
significantly more complicated (see the last paragraph of Section
\ref{btsec}). Our main approach to proving polynomial time
solvability and NP-hardness of special cases of \MiP is the use of
various reductions to/from the maximum weight independent set
problem.

Suppose we are given a pair of digraphs $D,H$ and a positive
integral cost $c_i(u)$ for each $u\in V(D)$ and $i\in V(H)$. The
{\em cost} of a homomorphism $f$ of $D$ to $H$ is $\sum_{u\in
V(D)}c_{f(u)}(u)$. For a fixed digraph $H$, the {\em minimum cost
homomorphism problem} \MiP is formulated as follows. For an input
digraph $D$ and costs $c_i(u)$ for each $u\in V(D)$ and $i\in
V(H)$, verify whether $HOM(D,H)\neq \emptyset$ and, if
$HOM(D,H)\neq \emptyset$, find a homomorphism in $HOM(D,H)$ of
minimum cost. The {\em maximum cost homomorphism problem} \MaP is
the same problem as MinHOMP($H$), but instead of minimization we
consider maximization. Let $M$ be an integral constant larger than
any cost $c_i(u)$, $u\in V(D),\ i\in V(H)$. Then the cost
$c'_i(u)=M-c_i(u)$ is positive for each $u\in V(D),\ i\in V(H)$.
Due to this transformation, the problems \MiP and \MaP are
equivalent.

For a digraph $G$, if $xy\in A(G)$, we say that $x$ {\em dominates}
$y$ and $y$ is {\em dominated} by $x$ (denoted by $x\dom y$). The
{\em outdegee} $d^+_G(x)$ ({\em indegree} $d_G^-(x)$) of a vertex
$x$ in $G$ is the number of vertices dominated by $x$ (that dominate
$x$). For sets $X,Y\subset V(G)$, $X\dom Y$ means that $x\dom y$ for
each $x\in X,\ y\in Y$, but no vertex of $Y$ dominates a vertex in
$X.$ A set $X\subseteq V(G)$ is {\em independent} if no vertex in
$X$ dominates a vertex in $X.$ A $k$-{\em cycle}, denoted by
$\vec{C}_k$, is a directed simple cycle with $k$ vertices. A digraph
$H$ is an {\em extension} of a digraph $D$ if $H$ can be obtained
from $D$ by replacing every vertex $x$ of $D$ with a set $S_x$ of
independent vertices  such that if $xy\in A(D)$ then $uv\in A(H)$
for each $u\in S_x,\ v\in S_y.$ For subsets $X,Y$ of $V(G)$,
$X\times Y=\{xy: x\in X,\ y\in Y\}.$

The {\em underlying graph} $U(G)$ of a digraph $G$ is the undirected
graph obtained from $G$ by disregarding all orientations and
deleting one edge in each pair of parallel edges. A digraph $G$ is
{\em connected} if $U(G)$ is connected. The {\em components} of $G$
are the subdigraphs of $G$ induced by the vertices of components of
$U(G)$. A digraph $G$ is {\em strongly connected} if there is a path
from $x$ to $y$ for every ordered pair of vertices $x,y\in V(G).$ A
{\em strong component} of $G$ is a maximal induced strongly
connected subdigraph of $G.$ A digraph $G'$ is the {\em dual} of a
digraph $G$ if $G'$ is obtained from $G$ by changing orientations of
all arcs.

The rest of the paper is organized as follows. In Section
\ref{homsec}, we give all polynomial time solvable cases of \MiP
when $H$ is semicomplete $k$-partite digraph, $k\ge 3$, or a
bipartite tournament. Section \ref{smdsec} is devoted to a full
dichotomy classification of the time complexity of \MiP when $H$ is
a semicomplete $k$-partite digraph, $k\ge 3$. A classification of
the same problem for $H$ being a bipartite tournament is proved in
Section \ref{btsec}.

\section{Polynomial Time Solvable Cases}\label{homsec}

The following definitions and theorem were given in \cite{gutinsd}.
The {\em homomorphic product}  of digraphs $D$ and $H$ is an
undirected graph $D\otimes H$ defined as follows: $V(D\otimes
H)=\{u_i:\ u\in V(D),\ i\in V(H)\}$, $E(D\otimes H)=\{u_iv_j:\ uv\in
A(D),\ ij\notin A(H)\}\cup \{u_iu_j: u\in V(D), i\neq j\in V(H)\}.$
Let $\mu=\max\{c_j(v):\ v\in V(D),\ j\in V(H)\}.$ We define the cost
of $u_i$, $c(u_i)=c_i(u)+\mu |V(D)|.$ For a set $X\subseteq
V(D\otimes H)$, we define $c(X)=\sum_{x\in X}c(x).$

\begin{theorem}\label{maint}
Let $D$ and $H$ be digraphs. Then there is a homomorphism of $D$
to $H$ if and only if the number of vertices in a largest
independent set of $D\otimes H$ equals $|V(D)|$. If $HOM(D,H)\neq
\emptyset$, then a homomorphism $h\in HOM(D,H)$ is of maximum cost
if and only if $I=\{x_{h(x)}:\ x\in V(D)\}$ is an independent set
of maximum cost.
\end{theorem}

Let $TT_k$ denote the acyclic tournament on $k$ vertices. In
\cite{gutinsd} we used Theorem \ref{maint} in order to prove the
following result:

\begin{theorem}\label{tt}
${\rm MWHOMP}(TT_k)$ and $MCHOMP(TT_k)$ are polynomial time
solvable.
\end{theorem}

Here we prove another, somewhat more involved, corollary of
Theorem \ref{maint}. A digraph $D$ is {\em transitive} if
$xy,yz\in A(D)$ implies $xz\in A(D)$ for all pairs $xy,yz$ of arcs
in $D.$ A digraph $D$ is the {\em transitive closure} of a digraph
$H$ if $V(D)=V(H)$ and $D$ is has the minimum number of arcs with
respect to the following properties: $D$ is transitive and
$A(H)\subseteq A(D)$. It is easy to see the uniqueness of the
transitive close of a digraph \cite{bang2000}. A graph is a {\em
comparability graph} if it has an orientation, which is
transitive. Let $k\ge 3$ and let $TT_k^-$ be a digraph obtained
from $TT_k$ by deleting the arc from the source to the sink.

\begin{theorem}\label{tt-}
${\rm MWHOMP}(TT_k^-)$ and $MCHOMP(TT_k^-)$ are polynomial time
solvable.
\end{theorem}
\pf  Let  $V(TT_k^-)=\{1,2,\ldots ,k\}$ and let $A(TT_k^-)=\{ ij :\
1 \leq i<j \leq k \ j-i<k-1 \}$. Observe that $HOM(D,H)= \emptyset$
unless $D$ is acyclic. Since we can verify that $D$ is acyclic in
polynomial time (for example, by deleting vertices of indegree 0),
we may assume that $D$ is acyclic. We will furthermore assume that
$D$ has no isolated vertices as these can be assigned an optimal
color greedily. Let $S=\{ s :\ d_D^-(s)=0 \}$ and let $T=\{ t :\
d_D^+(t)=0 \}$. Let $P(x,y)$ denote the set of paths from a vertex
$x$ to a vertex $y$ in $D.$ Let $D'$ be a digraph obtained from $D$
by adding the following set of arcs:

$$\{xy:\ xy\in V(D)\times V(D)- (A(D)\cup (S\times T)),\ P(x,y)\neq \emptyset\}.$$

Note that $D'$ is the transitive closure of $D$, where we delete
all arcs from $S$ to $T$ which are not present in $D$. One can
find $D'$ in polynomial time using the depth first search or
breadth first search \cite{cormen2001}. If $h \in HOM(D,TT_k^-)$
and $xy \in A(D')-A(D)$ then $x \not\in S$ or $y \not\in T$, which
implies that $h(x) \not=1$ or $h(y) \not=k$, which again implies
that $h(x)h(y) \in A(TT_k^-)$. This implies that
$HOM(D,TT_k^-)=HOM(D',TT_k^-)$.

Observe that for every $TT_k^-$-coloring $h$ of $D$ and each $s\in
S,\ t\in T,\ z\in V(D)-(S\cup T)$, we have $h(s)<k$, $h(t)>1$ and
$1<h(z)<k.$ Consider $$G=D'\otimes TT_k^--(\{s_k:\ s\in S\}\cup
\{t_1:\ t\in T\}\cup \{z_1,z_k:\ z\in V(D)-(S\cup T)\}.$$ By
Theorem \ref{maint} and the above observations, $HOM(D,H)\neq
\emptyset$ if and only if the number of vertices in a largest
independent set of $G$ equals $|V(D)|$. Moreover, a homomorphism
$h\in HOM(D',TT_k^-)$ is of maximum cost if and only if
$I=\{x_{h(x)}:\ x\in V(D)\}$ is an independent set of maximum
cost.

If we prove that $G$ is a comparability graph, we will be able to
find a largest independent set in $G$ and an independent set of
$G$ of maximum cost in polynomial time
\cite{grotschelADM21,kagarisI27}. So, it remains to find an
orientation of $G$, which is a transitive digraph.

Define an orientation $G^*$ of $G$ as follows: $A(G^*)=A_1 \cup
A_2\cup A_3$, where
  $$\begin{array}{l}A_1=\{ y_i z_j :\ yz \in A(D'),\ i\ge j
  \},
  A_2=\{ x_i x_j :\ x \in V(D),\ i>j \},\\
  A_3=\{ t_k s_1 :\ st \in A(D),\ s \in S, \ t \in T \}.\end{array}$$

Let $u_iv_j,v_jw_l\in A(G^*).$ We will prove that $u_iw_l\in
A(G^*).$ Consider three cases covering all possibilities without
loss of generality.

\2

 {\bf Case 1:} $u=v=w$. Since $u_iu_j,u_ju_l\in A_2$, we have
 $i>j>l$ and
$u_iu_l\in A_2.$

\2

{\bf Case 2:} $u=v\neq w.$ Then $u_iu_j\in A_2$, $u_jw_l\in A_1$
and, thus, $i>j\ge l$ and $uw\in A(D')$ implying $u_iw_l\in A_1.$

\2

 {\bf Case 3:} $|\{u,v,w\}|=3.$ Since $u_iv_j,v_jw_l\in A_1$, we have
 $uv,vw\in A(D')$ and thus $uw\in A(D').$ Also, $i\ge j\ge
 l.$ We conclude that $u_iw_l\in A_1.$
\qed

\begin{lemma}\label{lemext}
Suppose that \MiP is polynomial time solvable then, for each
extension $H'$ of $H$, MinHOMP($H'$) is also polynomial time
solvable.
\end{lemma}
\pf Recall that we can obtain $H'$ from $H$ by replacing every
vertex $i\in V(H)$ with a set $S_i$ of independent vertices.
Consider an $H'$-coloring $h'$ of an input digraph $D$. We can
reduce $h'$ into an $H$-coloring of $D$ as follows: if $h'(u)\in
S_i$, then $h(u)=i.$

Let $u\in V(D)$. Assign $\min\{c_j(u):\ j\in S_i\}$ to be  a new
cost $c_{i}(u)$ for each $i\in V(H)$. Observe that we can find an
optimal $H$-coloring $h$ of $D$ with the new costs in polynomial
time and transform $h$ into an optimal $H'$-coloring of $D$ with
the original costs using the obvious inverse of the reduction
described above.\qed

\2

In \cite{gutinsd}, we proved that \MiP is polynomial time solvable
when $H=\vec{C}_k$, $k\ge 2.$ Combining this results with Theorems
\ref{tt}, \ref{tt-} and Lemma \ref{lemext}, we immediately obtain
the following:

\begin{theorem}\label{poly} If $H$ is an extension of $TT_k$, $\vec{C}_k$ or $TT_k^-$ $(k\ge
3)$, then \MiP and \MaP are polynomial time solvable.
\end{theorem}

\begin{theorem}\label{bt} Let $H$ be an acyclic bipartite tournament. Then
\MiP and \MaP are polynomial time solvable.
\end{theorem}
\pf A pair of vertices in $H$ is called {\em similar} if they have
the same set of out-neighbors and the same set of in-neighbors;
$H$ is {\em simple} if it has no similar vertices. Let $V_1,V_2$
be the partite sets of $H$, let $D$ be an input digraph and let
$c_i(x)$ be the costs, $i\in V(H),\ x\in V(D).$ Observe that if
$D$ is not bipartite, then $HOM(D,H)=\emptyset$, so we may assume
that $D$ is bipartite. We can check whether $D$ is bipartite in
polynomial time using, e.g., the breadth first search
\cite{cormen2001}. Let $U_1,U_2$ be the partite sets of $D.$

To prove that we can find a minimum cost $H$-coloring of $D$ in
polynomial time, it suffices to show that we can find a minimum
cost $H$-coloring $f$ of $D$ such that $f(U_1)=V_1$ and
$f(U_2)=V_2.$ Indeed, if $D$ is connected, to find a minimum cost
$H$-coloring of $D$ we can choose from a minimum cost $H$-coloring
$f$ with $f(U_1)=V_1$ and $f(U_2)=V_2$ and a minimum cost
$H$-coloring $h$ of $D$ with $h(U_1)=V_2$ and $h(U_2)=V_1.$ If $D$
is not connected, we can find a minimum cost $H$-coloring of each
component of $D$ separately.

To force $f(U_1)=V_1$ and $f(U_2)=V_2$ for each $H$-coloring $f$,
it suffices to modify the costs such that it is too expensive to
assign any color from $V_j$ to a vertex in $U_{3-j}$, $j=1,2.$ Let
$M=|V(D)|\cdot \max\{c_i(x):\ i\in V(H),\ x\in V(D)\}+1$ and let
$c_i(x):=c_i(x)+M$ for each pair $x\in U_j$, $i\in V_{3-j},$
$j=1,2.$

We consider the following two cases.

\2

{\bf Case 1:} $H$ is simple. Observe that the vertices of $H$ can
be labeled $i_1,i_2,\ldots,i_p$ such that $i_k$ is the only vertex
of in-degree zero in $H-\{i_1,i_2,\ldots,i_{k-1}\}.$ Thus, $H$ is
a spanning subdigraph of $TT_p$ with vertices $i_1,i_2,\ldots,i_p$
($i_si_t\in A(TT_p)$ if and only if $s<t$). Observe that
$$\{f\in HOM(D,H): f(U_j)=V_j,\ j=1,2\}=\{f\in HOM(D,TT_p): f(U_j)=V_j,\
j=1,2\}.$$ Thus, to solve \MiP with the modified costs it suffices
to solve MinHOMP($TT_p$) with the same costs. We can solve the
latter in polynomial time by Theorem \ref{tt}.

\2

{\bf Case 2:} $H$ is not simple. Then $H$ is an extension of an
acyclic simple bipartite tournament. Thus, we are done by Case 1
and Lemma \ref{lemext}.\qed

\section{Classification for  semicomplete $k$-partite digraphs, $k\ge
3$}\label{smdsec}

The following lemma allows us to prove that \MaP and \MiP are
NP-hard when MaxHOMP($H'$) and MinHOMP($H'$) are NP-hard for an
induced subdigraph $H'$ of $H.$

\begin{lemma}\label{red}\cite{gutinsd}
Let $H'$ be an induced subdigraph of a digraph $H$. If
MaxHOMP($H'$) is NP-hard, then \MaP is also NP-hard.
\end{lemma}

The following lemma is the NP-hardness part of the main result in
\cite{gutinsd}.

\begin{lemma}\label{sd} Let $H$ be a semicomplete digraph containing
a cycle and let $H\not\in \{\vec{C}_2,\vec{C}_3\}$. Then \MiP and
\MaP are NP-hard.
\end{lemma}

The following lemma was proved in \cite{gutjahr1991}.

\begin{lemma}\label{cycle++}
Let $H_1$ be a digraph obtained from $\vec{C}_3$ by adding an
extra vertex dominated by two vertices of the cycle and let $H$ be
$H_1$ or its dual. Then ${\rm HOMP}(H)$ is NP-complete.
\end{lemma}

We need two more lemmas for our classification.

\begin{lemma}\label{ac}
Let $H'$ be given by $V(H')=\{1,2,3,4\},\
A(H')=\{12,23,34,14,24\}$ and let $H$ be $H'$ or its dual. Then
\MaP and \MiP are NP-hard.
\end{lemma}
\pf We reduce the maximum independence set problem (MISP) to
MinHOMP($H$). Let $G$ be an arbitrary graph. We construct a
digraph $D$ from $G$ as follows: every vertex of $G$ belongs to
$D$ and, for each pair $x,y$ of non-adjacent vertices of $G$, we
add to $D$ new vertices $u=u(x,y)$ and $v=v(x,y)$ together with
arcs $ux,uv,vy$. (No edge of $G$ is in $D$.) Let $n$ be the number
of vertices in $D.$ Let $x,y$ be a non-adjacent pair of vertices
in $G$ and let $u=u(x,y),\ v=v(x,y).$ We set
$c_3(x)=c_3(y)=c_i(u)=c_i(v)=1$ for $i=1,2,3$, $c_4(x)=c_4(y)=n+1$
and $c_j(x)=c_j(y)=c_4(u)=c_4(v)=n^2+n+1$ for $j=1,2.$

Consider a minimum cost $H$-coloring $h$ of $D.$ Let $x,y$ be a
pair of non-adjacent vertices in $G$ and let $u,v$ be the vertices
added to $D$ due to $x,y$ such that $ux,uv,vy$ are arcs of $D$.
Due to the values of the costs, $h$ can assign $x,y$ only colors 3
and 4 and $u,v$ only colors 1,2,3. The coloring can assign $u$
either 1 or 2. If $u$ is assigned 1, then $v,y,x$ must be assigned
2,3 and 4, respectively. If $u$ is assigned 2, then $v,y,x$ must
be assigned 3,4 and (3 or 4), respectively. In both cases, only
one of the vertices $x$ and $y$ can receive color 3. Since $h$ is
optimal, the maximum number of vertices in $D$ that it inherited
from $G$ must be assigned color 3. This number is the maximum
number of independent vertices in $G.$ Since MISP is NP-hard, so
is MinHOMP($H$).\qed

\begin{lemma}\label{cycle+++}
Let $H$ be given by $V(H)=\{1,2,3,4\},\ A(H)=\{12,23,31,34,41\}$.
Then \MaP and \MiP are NP-hard.
\end{lemma}
\pf
 We will reduce the maximum independent set problem to MinHOMP($H$). However
 before we do this we consider a digraph $D^{gadget}(u,v)$
 defined as follows:
$V(D^{gadget}(u,v))=\{x,y,u',u,v',v,c_1,c_2,\ldots ,c_{12}\}$ and
$$A(D^{gadget}(u,v))=\{xy,xc_1,yc_1,c_6u',u'u,c_{11}v',v'v,
c_1c_2,c_2c_3,c_3c_4,\ldots,c_{11}c_{12},c_{12}c_1\}$$

Observe that in any homomorphism $f$ of $D^{gadget}(u,v)$ to $H$
we must have $f(c_1)=1$, by the existence of $x$ and $y$. This
implies that $(f(c_1),f(c_2), \ldots ,f(c_{12}))$ has to coincide
with one of the following two sequences:

$$ (1,2,3,1,2,3,1,2,3,1,2,3) \mbox{ or }
(1,2,3,4,1,2,3,4,1,2,3,4).$$

If the first sequence is the actual one, then we have $f(c_6)=3$,
$f(u') \in \{1,4\}$, $f(u) \in \{1,2\}, f(c_{11})=2$, $f(v')=3$
and $f(v) \in \{1,4\}.$ If the second sequence is the actual one,
then we have $f(c_6)=2$, $f(u')=3$, $f(u) \in \{1,4\},
f(c_{11})=3$, $f(v') \in \{1,4\}$ and $f(v) \in \{1,2\}$. So in
both cases we can assign both of $u$ and $v$ color $1$.
Furthermore by choosing the right sequence we can color one of $u$
and $v$ with color $2$ and the other with color $1$.  However we
cannot assign color $2$ to both $u$ and $v$ in a homomorphism.

 Let $G$ be a graph. Construct a digraph $D$ as follows.
Start with $V(D)=V(G)$ and, for each edge $uv \in E(G)$, add a
distinct copy of $D^{gadget}(u,v)$ to $D$. Note that the vertices
in $V(G)$ form an independent set in $D$ and that
$|V(D)|=|V(G)|+16|E(G)|$.

Let all costs $c_i(t)=1$ for $t\in V(D)$ apart from $c_{1}(p)=2$
for all $p\in V(G).$ Clearly, a minimum cost $H$-coloring $h$ of
$D$ must aim at assigning as many vertices of $V(G)$ in $D$ a
color different from $1$. However, if $pq$ is an edge in $G$, by
the arguments above, $h$ cannot assign colors different from $1$
to both $p$ and $q$. However, $h$ can assign colors different from
$1$ to either $p$ or $q$ (or neither). Thus, a minimum cost
homomorphism of $D$ to $H$ corresponds to a maximum independent
set in $G$ and vise versa (the vertices of a maximum independent
set are assigned color $2$ and all other vertices in $V(G)$ are
assigned color $1$). \qed

\2

\begin{theorem}
Let $H$ be a semicomplete $k$-partite digraph, $k\ge 3$. If $H$ is
an extension of $TT_k$, $\vec{C}_3$ or $TT_p^-$ $(p\ge 4)$, then
\MiP and \MaP are polynomial time solvable. Otherwise, \MiP and
\MaP are NP-hard.
\end{theorem}
\pf We assume that P$\neq$NP as otherwise this theorem is of no
interest. Since $H$ is a semicomplete $k$-partite digraph, $k\ge
3$, if $H$ has a cycle, then there can be three possibilities for
the length of a shortest cycle in $H$: 2,3 or 4. Thus, we consider
four cases, the above three cases and the case when $H$ is
acyclic.

\2

{\bf Case 1:} $H$ has a 2-cycle $C$. Let $i,j$ be vertices of $C$.
The vertices $i,j$ together with a vertex from a partite set
different from those where $i,j$ belong to form a semicomplete
digraph with a 2-cycle. Thus, by Lemmas \ref{red} and \ref{sd},
\MiP and \MaP are NP-hard.

\2

{\bf Case 2:} A shortest cycle $C$ of $H$ has three vertices
$i,j,l$ ($ij,jl,li\in A(C)$). If $H$ has at least four partite
sets, then \MiP and \MaP can be shown be be NP-hard similarly to
Case 1. Assume that $H$ has three partite sets and that \MiP and
\MaP are polynomial time solvable.  Let $V_1$, $V_2$ and $V_3$ be
partite sets of $H$ such that $i\in V_1$, $j\in V_2$ and $l\in
V_3$. Let $s$ be a vertex outside $C$ and let $s\in V_1$. If $s$
is dominated by $j$ and $l$ or dominates $j$ and $l$, then \MiP
and \MaP are NP-hard by Lemmas \ref{red} and \ref{cycle++}, a
contradiction. If $j\dom s\dom k$, then \MiP and \MaP are NP-hard
by Lemmas \ref{red} and \ref{cycle+++}, a contradiction. Thus,
$k\dom s\dom j.$ Similar arguments show that $k\dom V_1\dom j.$
Let $p\in V_2-\{j\}.$ Similar arguments show that $p\dom V_1\dom
j$ and moreover $V_3\dom V_1\dom j.$ Again, similarly we can prove
that $V_3\dom V_1\dom V_2$, i.e., $H$ is an extension of
$\vec{C}_3$.

\2

{\bf Case 3:} A shortest cycle $C$ of $H$ has four vertices
$i,j,s,t$ ($i\dom j\dom s\dom t\dom i$). Since $C$ is a shortest
cycle, $i,s$ are belong to the same partite set, say $V_1$, and
$j,t$ belong to the same partite set, say $V_2$. Since $H$ is not
bipartite, there is a vertex $l$ belonging to a partite set
different from $V_1$ and $V_2.$ Since $H$ has no cycle of length 2
or 3, either $l$ dominates $V(C)$ or $V(C)$ dominates $l.$
Consider the first case $(l\dom V(C)$) as the second one can be
tackled similarly. Let $H'$ is the subdigraph of $H$ induced by
the vertices $l,i,j,s$. Observe now that \MiP and \MaP are NP-hard
by Lemmas \ref{red} and \ref{ac}.

\2

{\bf Case 4:} $H$ has no cycle. Assume that  \MiP and \MaP are
polynomial time solvable, but $H$ is not an extension of an acyclic
tournament. The last assumption implies that there is a pair of
nonadjacent vertices $i,j$ and a distinct vertex $l$ such that
$i\dom l\dom j.$ Let $s$ be vertex belonging to a partite set
different from the partite sets where $i$ and $l$ belong to. Without
loss of generality, assume that at least two vertices in the set
$\{i,j,l\}$ dominate $s.$ If all three vertices dominate $s$, then
by Lemmas \ref{red} and \ref{ac}, \MiP and \MaP are NP-hard, a
contradiction. Since $H$ is acyclic, we conclude that $\{i,l\}\dom
s\dom j.$ Let $V_1$ be the partite set of $i$ and $j.$ Similar
arguments show that for each vertex $t\in V(H)-V_1$, $i\dom t\dom
j.$ By considering a vertex $p\in V_1-\{i,j\}$ and using arguments
similar to the once applied above, we can show that either $p\dom
(V(H)-V_1)$ or $(V(H)-V_1)\dom p.$ This implies that we can
partition $V_1$ into $V_1'$ and $V_1''$ such that $V_1'\dom
(V(H)-V_1)\dom V_2''$. This structure of $H$ implies that there is
no pair $a,b$ of nonadjacent vertices in $V(D)-V_1$ such that $a\dom
c\dom b$ for some vertex $c\in V(H).$ Thus, the subdigraph $H-V_1$
is an extension of an acyclic tournament and, therefore, $H$ is an
extension of $TT_k^-.$\qed

\section{Classification for bipartite tournaments}\label{btsec}

The following lemma can be proved similarly to Lemma
\ref{cycle++}.

\begin{lemma}\label{cycleBT}
Let $H_1$ be given by $V(H_1)=\{1,2,3,4,5\},\
A(H_1)=\{12,23,34,41,15,35\}$ and let $H$ be $H_1$ or its dual.
Then \MaP and \MiP are NP-hard.
\end{lemma}

Now we can obtain a dichotomy classification for \MaP and \MiP
when $H$ is a bipartite tournament.

\begin{theorem}
Let $H$ be a bipartite tournament. If $H$ is acyclic or an
extension of a $4$-cycle, then \MiP and \MaP are polynomial time
solvable. Otherwise, \MiP and \MaP are NP-hard.
\end{theorem}

\pf If $H$ is an acyclic bipartite tournament or an extension of a
4-cycle, then \MiP is polynomial time solvable by Theorems
\ref{bt} and \ref{poly}. We may thus assume that $H$ has a cycle
$C$, but $H$ is not an extension of a cycle. We have to prove that
\MiP and \MaP are NP-hard.

Let $C$ be a shortest cycle of $H$. Since $H$ is a bipartite
tournament, we note that $|V(C)|=4$, so assume without loss of
generality that $C=w_1w_2w_3w_4w_1$, where $w_1,w_3$ belong to a
partite set $V_1$ of $H$ and $w_2,w_4$ belong to the other partite
set $V_2$ of $H$.

Observe that any vertex in $V_1$ must dominate either $w_2$ or
$w_4$ and must be dominated by the other vertex in $\{w_2,w_4\}$,
as otherwise we are done by Lemmas \ref{cycleBT} and \ref{red}.
Analogously any vertex in $V_2$ must dominate exactly one of the
vertices in $\{w_1,w_3\}$. Therefore, we may partition the
vertices in $H$ into the following four sets.

\begin{center}
$\begin{array}{ccc}
 W_1 = \{ v_1 \in V_1 :\  w_4 \dom v_1 \dom w_2 \} & \hspace{0.7cm} & W_2 = \{ v_2 \in V_2 :\  w_1 \dom v_2 \dom w_3 \} \\
 W_3 = \{ v_3 \in V_1 :\  w_2 \dom v_3 \dom w_4 \} &                & W_4 = \{ v_4 \in V_2 :\  w_3 \dom v_4 \dom w_1 \} \\
\end{array}$
\end{center}

\2

If $q_2q_1 \in A(H)$, where $q_j \in W_j$ for $j=1,2$ then we are
done by Lemma \ref{cycleBT} (consider the cycle $q_1 w_2 w_3 w_4
q_1$ and the vertex $q_2$ which dominates both $q_1$ and $w_3$).
Thus, $W_1 \dom W_2$ and analogously we obtain that $W_2 \dom W_3
\dom W_4 \dom W_1$, so $H$ is an extension  of a cycle, a
contradiction. \qed

\2

To find a complete dichotomy for the case of semicomplete
bipartite digraphs, one would need, among other things, to solve
an open problem from \cite{gutinDAM}: establish a dichotomy
classification for the complexity of \MiP when $H$ is a bipartite
(undirected) graph. Indeed, let $B$ be a semicomplete bipartite
digraph with partite sets $U,V$ and arc set $A(B)=A_1\cup A_2$,
where $A_1=U\times V$ and $A_2\subseteq V\times U$. Let $B'$ be a
bipartite graph with partite sets $U,V$ and edge set $E(B')=\{uv:\
vu\in A_2\}.$ Observe that MinHOMP($B$) is equivalent to
MinHOMP($B'$).

\2

\2

\noindent{\bf Acknowledgement} Research of Gutin and Rafiey was
supported in part by the IST Programme of the European Community,
under the PASCAL Network of Excellence, IST-2002-506778.

\end{document}